%% file: acl.tex
\documentclass[11pt]{article}

\usepackage[preprint]{acl}

\usepackage{times}
\usepackage{latexsym}

\usepackage[T1]{fontenc}
\usepackage[utf8]{inputenc}

\usepackage{amsmath}
\usepackage{microtype}

\usepackage{inconsolata}

\usepackage{graphicx}

\usepackage{algorithm}
\usepackage{algorithmic}

\usepackage{booktabs}
\usepackage{multirow}
\usepackage{makecell}
\usepackage{tabularx}
\usepackage{caption}
\usepackage{adjustbox}

\title{De-Anonymization at Scale via Tournament-Style Attribution}

\author{Lirui Zhang\\
  Beihang University \\
  \texttt{fysszlr@buaa.edu.cn} \\\And   
  Huishuai Zhang\thanks{Corresponding author.} \\
  Peking University \\
  \texttt{zhanghuishuai@pku.edu.cn} \\}

\begin{document}
\maketitle
\begin{abstract}

As LLMs rapidly advance and enter real-world use, their privacy implications are increasingly important. We study an authorship de-anonymization threat: using LLMs to link anonymous documents to their authors, potentially compromising settings such as double-blind peer review. 
We propose De-Anonymization at Scale (DAS), a large-language-model–based method for attributing authorship among tens of thousands of candidate texts. DAS uses a sequential progression strategy: it randomly partitions the candidate corpus into fixed-size groups, prompts an LLM to select the text most likely written by the same author as a query text, and iteratively re-queries the surviving candidates to produce a ranked top-k list. To make this practical at scale, DAS adds a dense-retrieval prefilter to shrink the search space and a majority-voting–style aggregation over multiple independent runs to improve robustness and ranking precision. Experiments on anonymized review data show DAS can recover same-author texts from pools of tens of thousands with accuracy well above chance, demonstrating a realistic privacy risk for anonymous platforms. On standard authorship benchmarks (Enron emails and blog posts), DAS also improves both accuracy and scalability over prior approaches, highlighting a new LLM-enabled de-anonymization vulnerability.

\end{abstract}

\input{introduction}

\input{related2}

\begin{figure*}[t]
    \centering
    \includegraphics[width=0.9\textwidth]{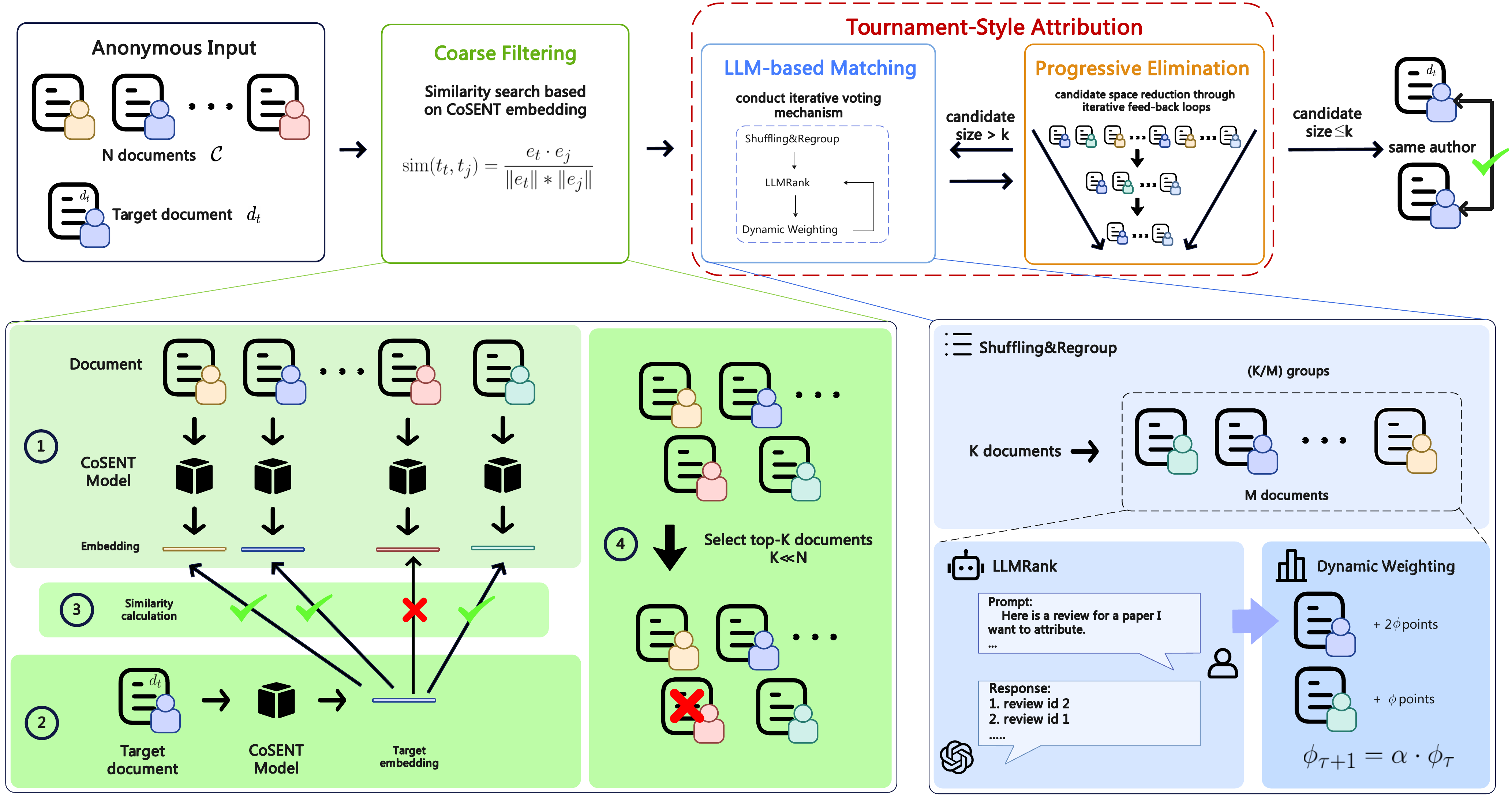}
    \caption{The DAS framework addresses the challenge of authorship attribution in large document corpora through a sequential progression strategy. The system operates in two core phases: 
(1) \textbf{Coarse Filtering} to narrow down candidate authors from thousands to a tractable subset, 
(2) \textbf{Tournament-Style Attribution (TSA)} that integrates LLMRank with dynamic weighting, multiple independent trials, and progressive elimination to iteratively reduce the candidate space and pinpoint the target author. 
This framework enables efficient analysis of long-form texts while maintaining computational feasibility at scale.}
    \label{fig:framework}
\end{figure*}

\section{Method}

This section presents the DAS framework for authorship attribution in large corpora. Section 3.1 formalizes the 
de-anonymization problem. Section 3.2 introduces DAS’s three-phase processing pipeline. Section 3.3 elaborates on the key algorithmic modules, including Coarse Filtering, LLM-based Matching, and Progressive Elimination.

\subsection{The De-Anonymization Problem }

The de-anonymization task in this paper aims to identify documents sharing the same author within a large corpus of anonymized texts. Formally, let $\mathcal{C}=\{d_i\}_{i=1}^N$ denote a collection of documents where each document $d_i=(t_i,a_i)$ consists of text content $t_i$ and hidden author identity $a_i\in \mathcal{A}$. Given a target document $d_t=(t_t,a_t)$, the task's objective is to obtain a subset of documents:
\begin{equation}
\mathcal{R}^* = \left\{d_j \in \mathcal{C} \setminus \{d_t\} \mid a_j = a_t\right\}.
\end{equation}
% We decompose this challenge into two computational subproblems:

% \textbf{Subproblem 1 (Coarse Filtering)}: 
%     Find a reduced set $S_K \subset \mathcal{C}$ containing the $K$-most probable candidates ($K \ll N$):
%     \begin{equation}
%     S_K = \mathop{\text{arg max}}\limits_{|S| = K} \sum_{d_j \in S} \text{sim}(t_t, t_j),
%     \end{equation}
%     where $\text{sim}(\cdot)$ measures stylistic similarity between documents.
    
% \textbf{Subproblem 2 (Precise Verification)}:
%     Determine the final $k$ document set $\hat{\mathcal{R}} \subset S_K$ using LLMs ($k \ll K$):
%     \begin{equation}
%     \hat{\mathcal{R}} = \mathop{\text{arg min}}\limits_{|\hat{\mathcal{R}}| = k} \sum_{d_j \in \hat{\mathcal{R}}} \mathop{\text{LLM}_\text{pref}}(t_t, t_j),
%     \end{equation}
%     where $\text{LLM}_\text{pref}(\cdot)$ indicates the model's preference rank between document pairs.

\subsection{De-Anonymization at Scale  Framework}

Our De-Anonymization at Scale (DAS) framework addresses the "needle in a haystack" challenge of authorship attribution through a dual-stage architecture. To maintain both computational feasibility and high precision, DAS decomposes the problem into two sequential objectives: \textbf{Stage 1 (Coarse Filtering)} rapidly narrows the search space from tens of thousands to hundreds of candidates, which keeps high recall rate and reduces the set of candidates significantly,   and when the search space is manageable, \textbf{Stage 2 (Tournament-Style  Attribution)} utilizes the deep reasoning capabilities of LLMs to pinpoint candidate texts of the same author. Notably, Stage 1 is optional and can be omitted when the candidate set is not large.

\subsubsection{Coarse Filtering} In DAS, we use vectorized retrieval to accomplish the Coarse Filtering mission, in order to address the efficiency bottleneck. It transforms the entire corpus  into a stylistic vector embedding. As detailed in Algorithm \ref{alg:VSR}, we use a CoSENT-based model~\citep{huang2024cosent,Text2vec}—a contrastive learning framework optimized for sentence embedding that normalizes cosine similarity scores to enhance semantic alignment—to generate fixed-dimension stylistic embeddings for every document. By computing the cosine similarity between the target document  and the corpus, we solve the first subproblem: identifying a reduced candidate set with top-$K$ maximized similarities.

\setcounter{algorithm}{0}
\begin{algorithm}[tb]
\caption{Coarse Filtering}
\label{alg:VSR}
\textbf{Input:} Target document $d_t$, Corpus $\mathcal{C}$, Size $K$ \\
\textbf{Output:} Filtered pool  $\mathcal{U}$
\begin{algorithmic}[1]
    \STATE{$e_t \gets \text{CoSENT}(t_t)$}
    \FOR{each document $d_j \in \mathcal{C}$}
        \STATE{$e_j \gets \text{CoSENT}(t_j), \quad s_j \gets \frac{e_t \cdot e_j}{\|e_t\| \cdot \|e_j\|}$}
        \STATE{Store $(d_j, s_j)$ in list $L$}
    \ENDFOR
    \STATE{Sort $L$ descending by $s_j$;} \\
    \STATE{$\mathcal{U} \gets L[1:K]$}
    \RETURN $\mathcal{U}$
\end{algorithmic}
\end{algorithm}

Importantly, this phase has to retain the ground-truth matches, i.e., high recall rates, when reducing the size of candidate set.

\subsubsection{Tournament-Style Attribution (TSA)}

Once the search space becomes manageable, DAS initiates Tournament-Style Attribution (Algorithm~\ref{alg:TSA}). This phase addresses the second subproblem: determining the final identity via a “survival of the fittest” style comparison. Next, following Algorithm~\ref{alg:TSA}, we detail its critical design components.

\textbf{Progressive Elimination.} Rather than processing  candidates simultaneously, the pool is partitioned into small groups. An LLM acts as a judge, performing deep linguistic analysis to rank candidates within each group.  By immediately removing lower-ranked candidates after each LLM query, it prevents the system from wasting resources on unlikely matches. This recursive reduction allows the LLM to focus on the most subtle stylistic differences among the ``finalists'', significantly increasing attribution accuracy while enjoying the tournament's efficiency. 

Specifically, in Algorithm~\ref{alg:TSA}, each group comparison retains 2 survivors and eliminates the remaining $l-2$ candidates.  For $\phi$ iterates as the size of the candidate set decreases to give higher scores to trusted documents. For each group, top-ranked documents receive $2\phi$ points, secondary choice $\phi$ points.

\begin{figure*}
    \centering
    \includegraphics[width=0.95\textwidth]{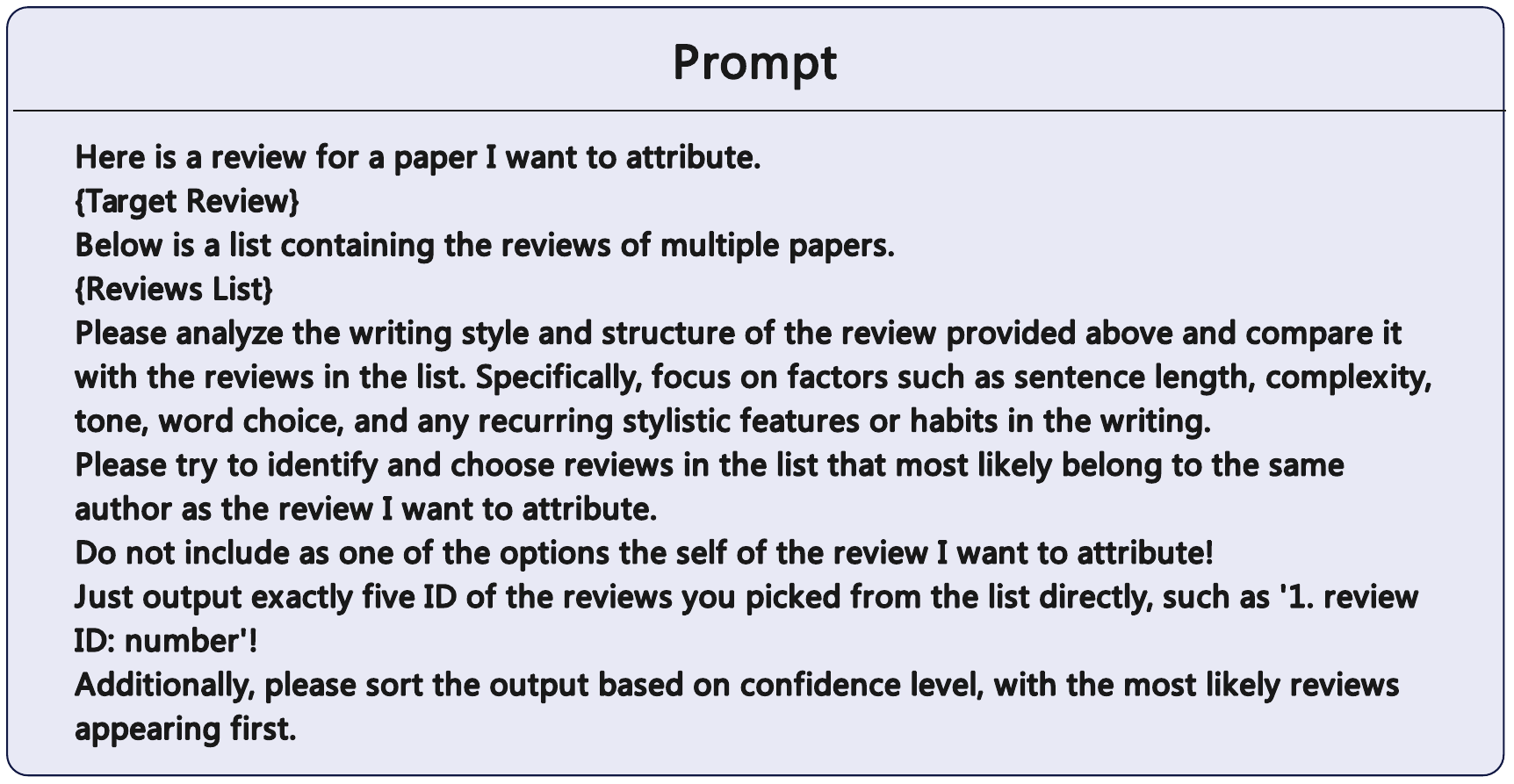}
    \caption{LLMRank prompt for paper review scenario.}
    \label{fig:prompt}
\end{figure*}

\textbf{LLMRank.} We randomly shuffle candidates and partition them into groups of size (l) to reduce ordering effects and fit the LLM context window. For each group $G_j=\{d_{j_1},\ldots,d_{j_l}\}$, we prompt the LLM with (i) the target text ($d_t$), (ii) the texts in ($G_j$), and (iii) instructions to compare stylistic cues, and ask it to rank candidates by likelihood of sharing the same author. The prompt template for paper review de-anonymization is provided in Figure~\ref{fig:prompt}.

\setcounter{algorithm}{1}
\begin{algorithm}
\caption{Tournament-Style Attribution (TSA)}
\label{alg:TSA}
\textbf{Input:} target text $t$, Filtered pool $\mathcal{U}$, Trials $m$, Scaling factor $\alpha$, group size $l$,  final $k$;  \\
\textbf{Output:} Ranked top-$k$ IDs $\hat{\mathcal{R}}$; \\
\begin{algorithmic}[1]
    \STATE Initialize global scores $\mathcal{S} \gets \{d: 0 \mid d \in \mathcal{U}\}$
    
    \FOR{Trial $t = 1$ \TO $m$}
        \STATE $\mathcal{O} \gets \text{shuffle}(\mathcal{U})$ \COMMENT{A new random trail} 
\\ $\phi \gets 1$ \COMMENT{A new random trail} 
        \WHILE{$|\mathcal{O} | > k$}
            \STATE Partition $\mathcal{O}$ into groups $\{G_1, \dots, G_p\}$ of size $l$ (last may be smaller),
            \STATE Let $\tilde{\mathcal{O}} \gets \emptyset$,
            
            \FOR{each group $G_j$}
                \STATE $\{ID_1, ID_2\} \gets \text{LLMRank}(d_t, G_j)$  \COMMENT{Top 2 IDs in $G_j$ that matches $d_t$ most}
                
                \STATE $\mathcal{S}[ID_1] \gets \mathcal{S}[ID_1] + 2\phi$\\ \COMMENT{Top rank reward}
                \STATE $\mathcal{S}[ID_2] \gets \mathcal{S}[ID_2] + \phi$ \\ \COMMENT{Runner-up reward}
                \STATE $\tilde{\mathcal{O}} \gets \tilde{\mathcal{O}}  \cup \{ID_1, ID_2\}$
            \ENDFOR
            
            \STATE $\mathcal{O} \gets \tilde{\mathcal{O}}$, \quad \COMMENT{Progressive Elimination} \\ 
            \STATE $\phi \gets \phi \cdot \alpha$ \quad \COMMENT{Dynamic Weighting}
        \ENDWHILE
    \ENDFOR
    
\STATE $\hat{\mathcal{R}} \leftarrow$ top-$k$ IDs in $\mathcal{U}$ sorted by $\mathcal{S}[\cdot]$ 
\RETURN $\hat{\mathcal{R}}$\end{algorithmic}
\end{algorithm}

\textbf{Dynamic Weighting.} To reinforce confidence in survivors, we use an exponential weighting mechanism . As the rounds progress and the candidate pool  shrinks, the scores awarded to survivors increase:
\begin{equation}
\phi_{\tau+1} = \alpha \cdot \phi_\tau,
\end{equation}
where $\phi_\tau$  is the current round and $\alpha$ is the acceleration rate. This ensures that authors who consistently \emph{win} their groups across multiple trials accumulate the highest global scores .

\textbf{Multiple Trials.} To be precise, Algorithm~\ref{alg:TSA} forms a \emph{multi-round survival tournament}. We run the tournament for ($m$) independent trials, each with a fresh random shuffle and regrouping of the candidate pool, which reduces sensitivity to any single partition and mitigates stochastic LLM behavior. In each trial, candidates that survive more rounds (and are selected more frequently within groups) accumulate higher scores. We then aggregate scores across trials (majority-voting style) and rank candidates by their total score, so texts that consistently match the query under different random groupings rise to the top while spurious matches are suppressed.

The score map  persists across all  trials. Even if a document is accidentally eliminated in one trial due to LLM noise, its consistent performance in other trials will keep its global ranking high.

So far, we have established the DAS framework and its underlying design philosophy. Next, we conduct experiments to validate its effectiveness.

\section{Experiments} 

In this section, we evaluate the effectiveness and practical implications of DAS. Section~4.1 studies real-world risk on anonymized peer-review data. Section~4.2 evaluates DAS on cross-domain benchmarks, including blogs and emails. Section~4.3 analyzes sensitivity across genres by tracking rank progression over rounds. Finally, Section~4.4 reports controlled ablations to quantify the contribution of each component.

\subsection{Evaluation on Anonymous Peer Reviews}\label{sec:anonyous-system}

To assess DAS in a realistic anonymous setting, we first evaluate it on anonymized conference peer-review data. This setup models an adversary attempting to compromise double-blind review by linking reviews of the same author via automated authorship analysis.

\subsubsection{Experiment Setup}

Specifically, we target on the following setup: Given a target annoymous review,  where the underlying reviewer identity is hidden, we apply DAS to retrieve other reviews likely written by the same reviewer.

\textbf{Dataset Construction.} We build a benchmark from ICLR 2023--2025 OpenReview records, covering 23,889 submissions and 147,367 anonymized reviews collected via the OpenReview API. Each instance includes submission metadata (e.g., title and abstract) and an anonymized review with a randomized ID. Table~\ref{tab:detail} summarizes key statistics. Given a query review

\textbf{Human Evaluation.} Since reviewer identities are not observable, we validate DAS with a human study. Nine participants provided 25 test cases and supplied one-shot judgments (see Appendix~\ref{app:human}) indicating which retrieved candidates were written by themselves. Because the study involves sensitive anonymous reviews from real-world reviewers, we minimized data collection: we stored no personal information and retained only aggregated voting outcomes over candidate ranges for analysis. This protocol prioritized and relied on mutual trust—participants provide honest labels, and we did not collect or retain identifiable data.

%For semantic representation learning, each review was transformed into a 768-dimensional vector through CoSENT-based encoding. We used \citet{Text2vec} as CoSENT base model.  Figure \ref{fig:distribute} visually shows the distribution of embeddings encoded. 

% \textbf{Human Evaluation (Real-World Trials).} Our evaluation includes a real-world human study: 9 participants contributed 25 unique test cases, where only one-time voting data were collected (details in Appendix~\ref{app:human}). Because the task operates on highly sensitive material, i.e., \emph{anonymous reviews from real reviewers}, the study design prioritized privacy and mutual trust. We did not retain any personal information during the experiment and preserved only the minimal outcomes, i.e., voting in ranges of candidates,  for analysis. This arrangement requires mutual trust: we trust participants not to submit random results, and participants trust that we do not collect or retain their personal data.

\begin{table}[htbp]
\centering
\begin{adjustbox}{width=0.95\columnwidth}
\footnotesize 
\begin{tabular}{@{}l ccc@{}}  
\toprule
Year & 2023 & 2024 & 2025 \\
\midrule
Paper number & 4955 & 7262 & 11672   \\
Review number & 27095& 44771 & 75501 \\
Mean Length & 3175 & 2562 & 2524 \\
\bottomrule
\end{tabular}
\end{adjustbox}
\caption{Statistical characteristics of the review dataset.}
\label{tab:detail}
\end{table}

% \begin{figure}[htbp] 
%     \centering
%     \includegraphics[width=1\linewidth]{fig/distribute.png}
%     \caption{Review embedding distribution per year. The dimensionality reduction of review embeddings into a two-dimensional latent space revealed consistent cluster alignments, evidencing similar distribution annually.}
%     \label{fig:distribute}
% \end{figure}

\textbf{Implementation Details of DAS.}  Given a target review $r_t$, we first apply coarse filtering to retrieve the \emph{top-$2000$} candidates. We then run the fine-grained stage for \emph{$m=5$} independent trials: in each trial, candidates are randomly grouped into clusters of \emph{$l=10$} reviews and compared by an LLM (Gemini-2.0-flash unless noted), using exponential round weighting with \emph{$\alpha=5$}. The tournament stops once \emph{$k\leq 20$} candidates remain, and we aggregate scores across trials to produce the final ranking. All experiments run on a single server with 4 vCPUs (Xeon Platinum 8269CY) and 8GB RAM. Each trial uses approximately 6,000 tokens (about \textasciitilde\$1 USD) and, with parallelization, finishes in 2\textasciitilde3 minutes.

\textbf{Evaluation Metric and Baseline.} We use \textbf{Rank@k} to evaluate the  effectiveness of the de-anonymization task, defined as the frequency that the top-$k$ list contains at least one same-author document.

As baselines, repeated human trials are infeasible due to reviewer availability and our privacy protocol, where we retain only voting outcomes, not full interaction logs. We therefore use \textbf{random guessing} as the primary baseline. For a candidate pool of size $N$, with $m$ same-author documents for a target review and cutoff $k$, the chance of retrieving at least one same-author document by random selection is:
\begin{equation}
\text{Rank@k}(\text{random}) = 1 - \frac{\binom{N-m}{k}}{\binom{N}{k}}.
\end{equation}
We report Rank@k relative to this baseline and show that DAS performs substantially better than random guessing. We include additional algorithmic baselines in the open-benchmark experiments that follow.

%We derived the theoretical expectation of random guessing through combinatorial mathematics: given a candidate pool containing $N$ documents, assuming each target document has $m$ same-author documents, the probability of selecting at least one same-author document when randomly choosing $k$ candidates is formulated as $\text{Rank@k} = 1 - \binom{N-m}{k}/\binom{N}{k}$. Through statistical computation, we obtained the rank and miss values for this random baseline, demonstrating the substantial performance enhancement of the DAS method compared to unordered guessing.

%\textbf{Evaluation Metrics} The evaluation framework employed two principal metrics. The Rank@k metric measures retrieval effectiveness by calculating the probability that at least one same-author document appears in the top-k ranked candidates. Precision@k evaluates ranking quality by computing the proportion of correctly identified same-author documents within the top-k results.

\subsubsection{Experiment Results}

Table~\ref{tab:real-scenario-review} shows that DAS is effective on large-scale anonymized reviews, achieving 28\% Rank@5 about a 1,000-fold improvement over random guessing. In practical terms, an attacker could reach roughly a 44\% success rate if inspecting only the top-20 candidates, indicating a substantive anonymity risk. Although the ICLR setting is challenging due to sparse stylistic cues and few same-author samples per reviewer, DAS remains robust, underscoring an emerging LLM-enabled privacy threat.

%Table \ref{tab:real-scenario-review} shows the DAS method's effectiveness in large-scale anonymized review datasets: achieving 28\% Rank@5 accuracy with a 1,000-fold improvement over random baselines, revealing substantial privacy risks - attackers could compromise anonymity with approximately half probability by inspecting just 20 candidate reviews. Despite limitations in the ICLR dataset from sparse textual features and limited same-author samples, the results validate the method's robustness when sufficient stylistic signals exist. These findings highlight emerging privacy threats enabled by LLMs.

\begin{table}[htbp]
\centering
\begin{adjustbox}{width=0.95\columnwidth}
\footnotesize 
\begin{tabular}{l*{5}{c}}
\toprule
& Rank@5 & Rank@10 & Rank@15 & Rank@20 & Miss \\
\midrule
DAS & 28\% & 40\% & 44\% & 44\% & 56\%   \\
Random & 0.03\% & 0.07\% & 0.10\% & 0.13\% & 99.88\% \\
\bottomrule
\end{tabular}
\end{adjustbox}
\caption{Real scenario experiments on ICLR anonymous reviewing systems.}
\label{tab:real-scenario-review}
\end{table}

\subsection{Evaluation on Existing Benchmarks} \label{sec:cross-domain}

Besides evaluating the emerging privacy risk  on anonymous peer review systems, we also conduct experiments on existing benchmarks that are widely used for distinguishing writing styles: blogs and emails. This test examines the capability of DAS on non-academic settings and supports broader claims about LLM-enabled de-anonymization.

\subsubsection{Experiment Setup}

\textbf{Datasets \& Baseline.} We use two public corpora. (1) \textit{Blog Authorship Corpus}~\citep{Schler2006EffectsOA}, containing posts from 19,320 bloggers; we sample 1,500 authors, each with 10 posts. (2) \textit{Enron Emails}~\citep{klimt2004enron}, a large email collection; following \citet{Huang2024CanLL}, we keep 174 authors with 50 emails each. As an algorithmic baseline, we report results using AIDBench~\citep{wen2024aidbench}, a recent framework for evaluating LLM-based authorship attribution under diverse conditions.

\begin{table*}[htbp]
\centering
\footnotesize 
\begin{tabular}{l*{6}{c}}
\toprule
\multirow{2}{*}{} & 
\multicolumn{3}{c}{2 Authors} & 
\multicolumn{3}{c}{5 Authors} \\
\cmidrule(lr){2-4} \cmidrule(lr){5-7}
& Rank@1 & Rank@3 & Rank@5 & Rank@1 & Rank@3 & Rank@5 \\
\midrule
DAS & 95.0 & 100.0 & 100.0 & 75.0 & 85.0 & 85.0\\
AIDBench & 66.7 & 93.3 &  93.3 & 66.7 & 86.7 & 86.7\\
\midrule
& Prec@1 & Prec@3 & Prec@5 & Prec@1 & Prec@3 & Prec@5 \\
\midrule
DAS & 95.0 & 85.0 & 76.0 & 75.0 & 66.7 & 60.0\\
AIDBench & 66.7 & 73.3 & 70.0 & 66.7 & 60.0 & 56.7\\
\bottomrule
\end{tabular}
\caption{Evaluation of the \emph{one-to-many}  author identification on the Blog Authorship Corpus (with Qwen1.5-72B-Chat model).}
\label{tab:one-to-many-blog}
\end{table*}

\begin{table*}[htbp]
\centering
\footnotesize 
\begin{tabular}{l*{6}{c}}
\toprule
\multirow{2}{*}{} & 
\multicolumn{3}{c}{2 Authors} & 
\multicolumn{3}{c}{5 Authors} \\
\cmidrule(lr){2-4} \cmidrule(lr){5-7}
& Rank@1 & Rank@3 & Rank@5 & Rank@1 & Rank@3 & Rank@5 \\
\midrule
DAS & 85.0 & 100.0 & 100.0 & 80.0 & 95.0 & 100.0\\
AIDBench & 66.7 & 93.3 & 100.0 & 66.7 & 73.3 & 80.0\\
\midrule
& Prec@1 & Prec@3 & Prec@5 & Prec@1 & Prec@3 & Prec@5 \\
\midrule
DAS & 85.0 & 83.3 & 77.0 & 80.0 & 71.7 & 72.0\\
AIDBench & 66.7 & 68.9 & 66.0 & 66.7 & 53.3 & 49.3\\
\bottomrule
\end{tabular}
\caption{Evaluation of the \emph{one-to-many} author identification on the Enron Email dataset (with Qwen1.5-72B-Chat model). }
\label{tab:one-to-many-email}
\end{table*}

\textbf{Experimental Design.} We run two complementary settings: (1) a \textit{one-to-many} test, where each query is mixed with a controlled number of distractors, and (2) a \textit{in-the-wild} test, where an adversary attempts to identify an author from an entire corpus rather than a pre-filtered candidate pool. Both follow the DAS pipeline with domain-specific prompts: for blogs, we emphasize narrative voice and colloquial phrasing; for emails, we focus on greetings, sign-offs, and topic transitions. Each configuration uses five independent trials with randomized candidate ordering. For consistency with the baseline, we use Qwen1.5-72B-Chat for the one-to-many tests, while using Gemini-2.0-flash for he in-the-wild tests.

\textbf{Evaluation Metrics.} We report two metrics. \textbf{Rank@k} measures retrieval success, i.e., the fraction of queries for which the top-\(k\) list contains at least one same-author document. \textbf{Precision@k} measures ranking quality, i.e., the proportion of same-author documents among the top-\(k\) results.

\subsubsection{Experiment Results}

\paragraph{One-to-many Test.}
Tables~\ref{tab:one-to-many-blog} and \ref{tab:one-to-many-email} show that DAS consistently outperforms the AIDBench baseline across both benchmarks. On blogs, DAS improves Rank@1 by up to 28\% in the two-author setting and maintains 75\% Precision@k in five-author identification. On Enron emails, DAS also yields higher accuracy, reaching 80\% precision in the five-author setting. Overall, DAS degrades more gracefully as the candidate pool grows, with substantially smaller performance drops than the baseline.

\paragraph{In-the-wild Test.} 
Table~\ref{tab:real-scenario-email} further confirms DAS in realistic search over the whole corpus. On blogs, Rank@k increases from 74\% at \(k{=}5\) to 94\% at \(k{=}20\), with only 6\% misses, suggesting the iterative tournament effectively captures consistent narrative cues. On Enron emails, where texts are shorter and stylistically diverse, DAS achieves 74\% Rank@5 and reaches 88\% by \(k{=}20\) (12\% misses). Across both datasets, multi-trial aggregation improves stability and mitigates occasional LLM errors.

\subsection{Ablation Studies on Progressive Elimination}

To isolate the contribution of progressive elimination to DAS's performance, we conducted controlled ablation experiments on the Blog Authorship Corpus. We define a variant DAS-PE that disables the progressive elimination mechanism: instead of iteratively retaining top candidates and shrinking the pool, DAS-PE processes all candidates in a single round of grouping and scoring without iterative reduction. All other components remain identical to the full DAS framework. We evaluate both the full DAS and DAS-PE using the same metrics to ensure direct comparability.

\paragraph{Experiment Results}

Table \ref{tab:abtest} demonstrates the critical role of progressive elimination through controlled experiments. Disabling progressive elimination (see variant DAS-PE) caused 34\% relative Rank@5 performance drop and doubled miss rates on blog data, with accuracy stagnating at 68\% (vs. 94\% full model). The complete system achieved noise suppression through iterative filtering, whereas DAS-PE demonstrated higher susceptibility to bias accumulation due to limited candidate screening. The results show the dual role of progressive elimination as a computational accelerator and a signal purifier.

\subsection{Ablation Studies on Different Models}

To further validate the robustness of DAS framework across different LLMs, we conducted additional experiments using two other popular language models DeepSeek-R1 and Claude-3.5-Sonnet. This extension aimed to assess the framework's adaptability to different models and reasoning approaches. The evaluation employed identical experimental parameters as previous tests (see Section \ref{sec:cross-domain}). 

\paragraph{Experiment Results}

We evaluate the one-to-many author identification task on the Research Paper dataset. 
As shown in Table \ref{tab:extended}, Claude-3.5-Sonnet attained perfect 100\% Rank@1 accuracy in two-author scenarios, while DeepSeek-R1 achieved 96.3\%, both achieving excellent results. %Other precision metrics also revealed Claude's superior performance. %Notably, the explicit reasoning mechanism in DeepSeek-R1 showed reduced effectiveness compared to Claude, suggesting potential limitations of CoT approaches in capturing nuanced authorial patterns.
%Both models maintained 100\% recall across conditions, confirming DAS's ability to leverage different LLMs' strengths while ensuring robust author identification. This model-agnostic performance highlights the framework's practical viability in diverse deployment environments.
These results confirm DAS's ability to leverage different LLMs' strengths while ensuring robust author identification. This model-agnostic performance highlights the framework's practical viability in diverse deployment environments.

\begin{table}[htbp]
\centering
\begin{adjustbox}{width=0.95\columnwidth}
\footnotesize 
\begin{tabular}{l*{5}{c}}
\toprule
& Rank@5 & Rank@10 & Rank@15 & Rank@20 & Miss \\
\midrule
DAS & 74\% & 88\% & 92\% & 94\% & 6\%   \\
DAS-PE & 40\% & 60\% & 66\% & 68\% & 32\%   \\
\bottomrule
\end{tabular}
\end{adjustbox}
\caption{Ablation test on blog dataset}
\label{tab:abtest}
\end{table}

% \subsubsection{Experiment Results}

% \paragraph{One-to-many Test}
% Tables \ref{tab:one-to-many-blog} and \ref{tab:one-to-many-email} demonstrate DAS's superior multi-author identification across domains. In blog analysis, DAS achieved 28\% higher Rank@1 accuracy than baseline for two-author scenarios, maintaining 75\% precision in five-author tasks. For Enron emails, DAS consistently outperformed AIDBench with 80\% precision in complex five-author identification. The framework's hierarchical design exhibited consistent performance advantages, particularly in scaled search scenarios where it showed 45\% lower error rate degradation compared to baseline approaches.

% \paragraph{In-the-wild Test}
% Table \ref{tab:real-scenario-email} demonstrates DAS's cross-domain effectiveness. For blogs, accuracy increased from 74\% (Rank@5) to 94\% (Rank@20) with 6\% misses, showing iterative scoring successfully captured narrative patterns. In emails with stylistic diversity, DAS achieved 74\% Rank@5 accuracy, stabilizing at 88\% (12\% misses), proving adaptability to short-text features. Multi-round trials with majority voting mitigate errors, validating consistency across fragmented business correspondence characteristics.

\begin{table}[htbp]
\centering
\begin{adjustbox}{width=0.95\columnwidth}
\footnotesize 
\begin{tabular}{l*{5}{c}}
\toprule
& Rank@5 & Rank@10 & Rank@15 & Rank@20 & Miss \\
\midrule
blog & 74\% & 88\% & 92\% & 94\% & 6\%   \\
email & 74\% & 86\% & 88\% & 88\% & 12\%   \\
\bottomrule
\end{tabular}
\end{adjustbox}
\caption{Evaluation of the \emph{in-the-wild} tests on Blog Authorship Corpus and Enron Emails.}
\label{tab:real-scenario-email}
\end{table}

\section{Conclusion}

In this paper, we propose De-Anonymization at Scale (DAS), a progressive framework that enables scalable authorship matching with LLMs. Real-scenario evaluations show that DAS can recover same-author texts in anonymized peer reviews (44\% Rank@20), blogs (94\% Rank@20), and emails (88\% Rank@20), exposing practical vulnerabilities in anonymous systems, where an attacker may succeed by inspecting only a small shortlist of candidates.

Looking ahead, methodologically, hybrid systems that combine LLM-based comparisons with classical stylometric signals may further strengthen attribution. More broadly, our results motivate standardized anonymity benchmarks and updated policies for deploying LLMs in privacy-sensitive settings, as well as closer collaboration between machine learning and privacy research to better protect anonymous communication in the LLM era.

\section{Limitations}

Our study demonstrates promising results, but has several limitations. First, our experiments focus mainly on Gemini-2.0-flash and Qwen1.5-72B-Chat, and may not fully reflect the behavior of other strong LLM architectures. Second, our prompt design is relatively limited, which may under-exploit LLMs’ capacity for fine-grained stylometric analysis.

\section{Ethical Considerations}

\textbf{Purpose and dual-use risk.} This work is a \emph{risk demonstration}: it shows that large language models (LLMs) can de-anonymize texts in anonymity-critical systems (e.g., double-blind peer review) by exploiting stylometric signals at scale. Our intent is to show actionable privacy risks so platforms and venues can assess and strengthen defenses, not to target individuals. Accordingly, our evaluation mirrors realistic conditions in anonymous ecosystems (open-set attribution, large candidate pools, and no pre-built author profiles).

\textbf{Data sources and scope.} All peer-review texts analyzed in this paper come from publicly accessible, \emph{anonymized} OpenReview records (ICLR 2023--2025). We use these texts only for evaluation and do not train or fine-tune any models on review content. DAS relies on retrieval and LLM prompting rather than supervised training on the review corpus, reducing risks of memorization or data leakage from these sources.

\textbf{Human subjects and privacy.} We complement offline evaluation with a small validation study (9 participants; 25 test cases) designed to protect participant privacy. We collect only the information needed to compute metrics and store no personally identifying information. Any potentially identifying inputs (e.g., paper identifiers) are hashed or deleted, and we retain only anonymized, aggregated voting outcomes (see Appendix~\ref{app:human}). The study design follows principles of notice, purpose limitation, data minimization, and storage limitation.

\textbf{Safeguards and non-targeting.} Because the study involves sensitive real-world material (anonymous reviews), we report only aggregated results and do not release per-document or per-user attributions. We also refrain from attempting to de-anonymize ongoing review threads outside controlled evaluation.

\textbf{Responsible communication.} To reduce misuse, we provide sufficient methodological detail for reproducibility while omitting operational instructions that would materially lower the barrier to abuse. We encourage venues to consider countermeasures such as stylometric audits, reviewer-style obfuscation tools, and clearer policies governing LLM use in privacy-sensitive workflows.

\textbf{Use of AI assistants.} We used LLMs only for language polishing (grammar and clarity). They were not used for code generation, data analysis, experimental design, or substantive content creation. All methods, results, and conclusions are the product of human work.

\section*{Acknowledgement}
The authors sincerely thank the volunteers, who are kept anonymous for safety reasons, for evaluating the system and offering invaluable real-world feedback. This work would have been substantially weaker without their contributions. The authors would also like to thank Zichen Wen and Dadi Guo for their contributions to the preparation of AIDBench and for their assistance in conducting the benchmark evaluation. The work was  supported by National Natural Science Foundation of China (Grant No. 62576015),  Beijing Major Science and Technology Project (Z251100008425004) and Beijing Natural Science Foundation (L253001).

\bibliography{privacy}

\newpage
\appendix

\noindent{\Large \textbf{Appendices}}

\section{Platform Functionalities} \label{app:human}

\begin{figure}[htbp]
  \centering
  \begin{minipage}[t]{0.48\textwidth}
    \includegraphics[width=\linewidth]{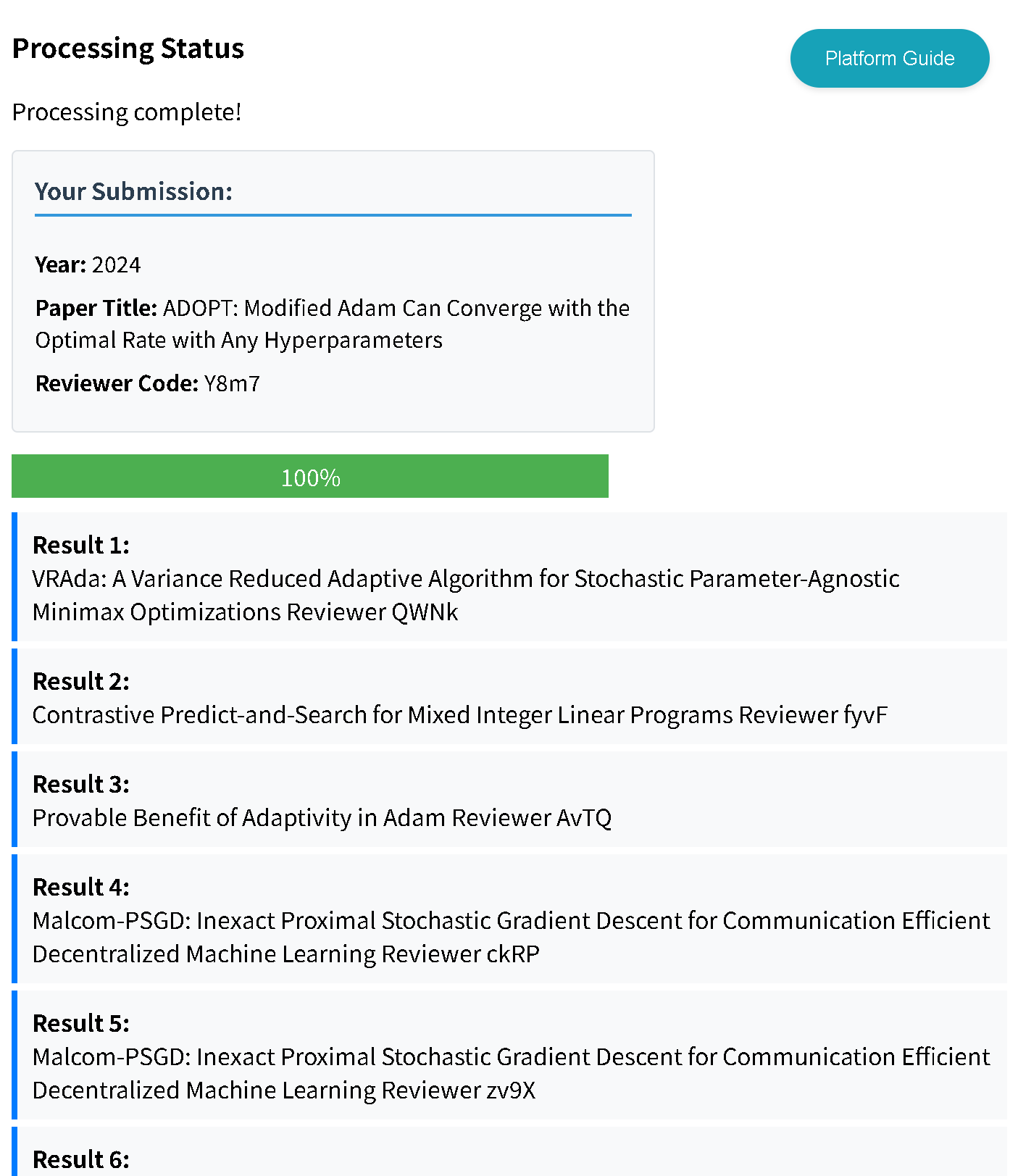}
  \end{minipage}
  \hfill 
  \begin{minipage}[t]{0.48\textwidth}
    \includegraphics[width=\linewidth]{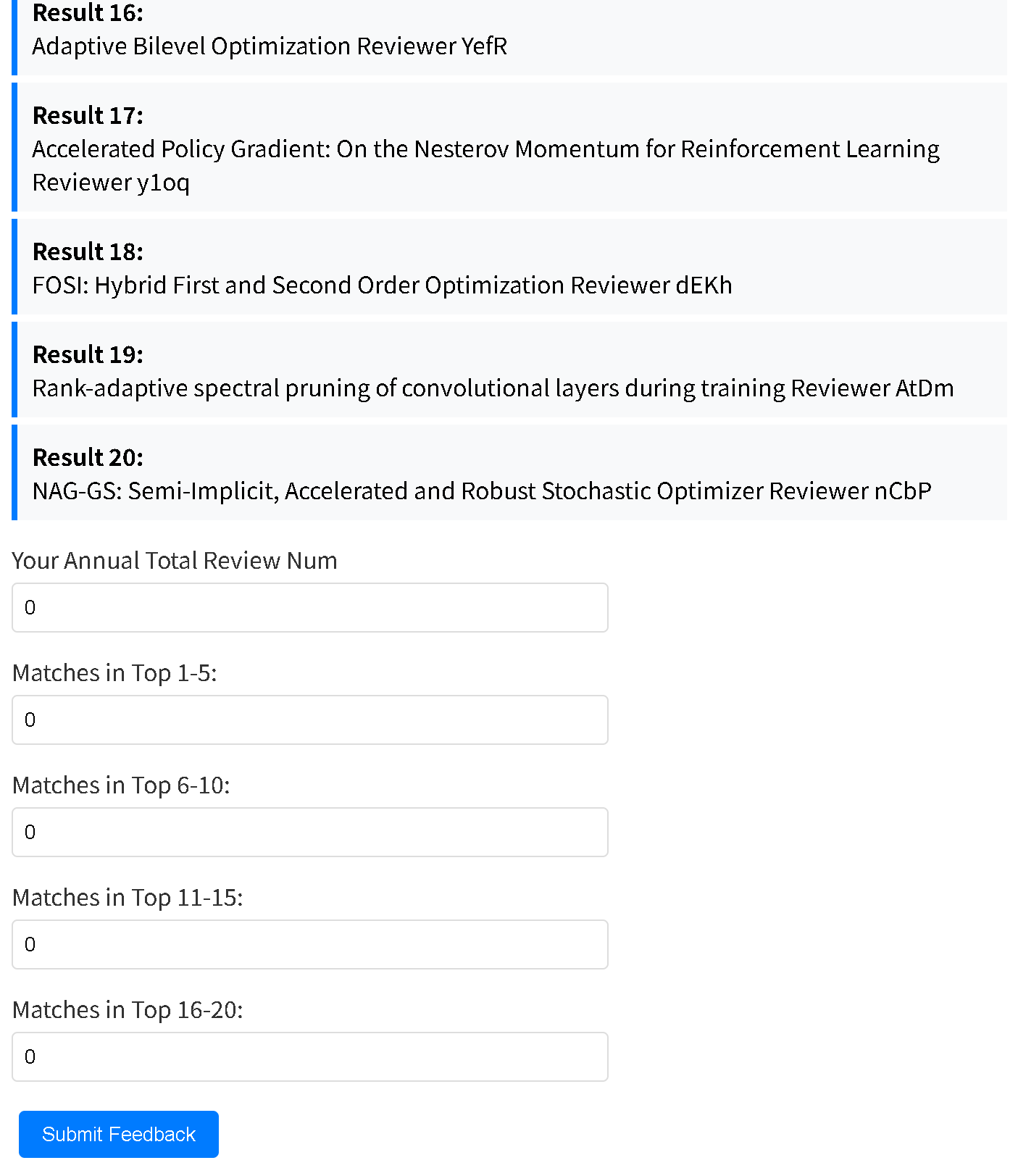}
  \end{minipage}
  \caption{User interface of our platform}
  \label{fig:platform}
\end{figure}

The web platform’s workflow centers on processing anonymous review analysis tasks in real‐world scenarios while providing a fluent user experience for our volunteer participants, who were not paid for their involvement. The process is presented in Figure~\ref{fig:platform} with a random query example.

First, participants submit a query specifying the year, paper title, and review code. That request enters a background task queue, which uses the DAS framework to retrieve the 20 most similar reviews and presents them as a ranked list from most to least likely. 

Next, participants report both the total number of reviews they authored that year and how many of those appear within each ranking interval. 

Finally, with these inputs, the system computes the necessary statistics. Under the hood, this workflow incorporates multiple technical components including asynchronous task processing, real-time status updates, dynamic result visualization, and privacy protection, ensuring efficient and safe evaluation.

We want to emphasize that this platform enabled researchers to evaluate the deanonymization risks inherent in anonymous review systems while strictly adhering to ethical guidelines and privacy protections.% We disclosed all information related to the project to the subjects before the experiment.

Before participants use the platform, we explicitly inform users of the platform's functionalities through pop-up notifications, including the purposes, types of information collected, and privacy protection measures. Throughout the entire interaction process, we ensure that users remain completely anonymous, preventing potential breaches in the anonymity system's privacy protections. As shown in Table \ref{tab:data-protocol}, the validation platform was designed in compliance with GDPR and COPPA regulations, incorporating multiple privacy safeguards.

\begin{table}[htbp]
\centering
\begin{adjustbox}{width=0.95\columnwidth}
\footnotesize 
\begin{tabular}{llll}
\toprule
Data Type & Description & Stored? & Protection Measures \\
\midrule
User Feedback & Rage hit statistics & Yes & Anonymization \\
User Inputs & Year, paper title, review code & No & Hash, Automatic deletion \\
System Logs & Task status information & No & Automatic deletion \\
\bottomrule
\end{tabular}
\end{adjustbox}
\caption{Data handling protocol for human subject interactions}
\label{tab:data-protocol}
\end{table}

%The functional workflow of this platform primarily revolves around processing anonymous review analysis tasks in real-world scenarios and facilitating user interaction. The process begins with participants submitting query requests containing the year, paper title, and review code. This triggers a background task queue that utilizes DAS framework to obtain the top 20 most similar reviews, which are displayed as a ranked list from the most possible to less possible. Subsequently, participants are required to input both the total number of  the reviews they wrote in that year and the number of matches within specified ranking intervals to calculate necessary statistics. This integrated workflow incorporates multiple technical components including asynchronous task processing, real-time status updates, dynamic result visualization, and systematic feedback data collection, ensuring efficient task execution and continuous method optimization. % validate the methodology's accuracy

\section{Evaluation on Research Paper Dataset}

To further investigate DAS's capability in identifying authorship of formal academic writing, we conducted comprehensive experiments on a large-scale research paper dataset. This evaluation extends our analysis to scholarly articles while addressing multi-author attribution challenges inherent in academic publications.

\subsection{Experiment Setup}

\paragraph{Benchmark Dataset \& Baseline} The research paper dataset~\citep{wen2024aidbench} comprised titles, abstracts and introductions from 138,652 arXiv papers under the CS.LG tag (the field of machine learning in the computer science domain) published between 2019 and 2024. After removing duplicate entries and authors with fewer than ten papers, the dataset included 24,095 papers from 1,500 authors, ensuring that each author had at least ten papers. As detailed in Section \ref{sec:cross-domain}, we implemented AIDBench~\citep{wen2024aidbench} as the baseline method.

\paragraph{Experiment Design}

Following the methodology established in Sections \ref{sec:anonyous-system} and \ref{sec:cross-domain}, we performed both one-to-many and real-scenario tests on the research paper dataset. The one-to-many test utilized Qwen1.5-72b-chat while the real-scenario test employed Gemini-2.0-flash. A successful identification was registered if the target author appeared in the author list of any retrieved result. This lenient evaluation criterion reflected real-world attack scenarios where attackers might only need to associate a document with one of its actual authors.

 \subsection{Experiment Results}

\paragraph{One-to-many Test}

The experimental results demonstrated that the DAS framework exhibited significant advantages in research paper authorship attribution tasks. As shown in Table \ref{tab:one-to-many-paper}, in two-author scenarios, DAS achieved a Rank@1 accuracy of 93.3\%, surpassing baseline methods by 23.3 percentage points. In the more challenging five-author scenarios, DAS attained a Rank@1 score of 60\% - significantly higher than the baseline. Regarding precision metrics, DAS obtained 48.9\% Precision@3 in five-author scenarios, representing a 38.9 percentage point improvement. This performance advantage proves particularly pronounced in long academic texts, likely stemming from DAS's proficiency in capturing discipline-specific terminological patterns, argumentative structures, and other deep stylistic features inherent to scholarly writing conventions.

\begin{table*}[htbp]
\centering
\footnotesize 
\begin{tabular}{l*{6}{c}}
\toprule
\multirow{2}{*}{} & 
\multicolumn{3}{c}{2 Authors} & 
\multicolumn{3}{c}{5 Authors} \\
\cmidrule(lr){2-4} \cmidrule(lr){5-7}
& Rank@1 & Rank@3 & Rank@5 & Rank@1 & Rank@3 & Rank@5 \\
\midrule
DAS & 93.3 & 93.3 & 100.0 & 60.0 & 80.0 & 86.7\\
AIDBench & 70.0 & 76.7 & 86.7 & 13.3 & 23.3 & 33.3\\
\midrule
& Prec@1 & Prec@3 & Prec@5 & Prec@1 & Prec@3 & Prec@5 \\
\midrule
DAS & 93.3 & 68.9 & 70.7 & 60.0 & 48.9 & 41.3\\
AIDBench & 70.0 & 51.1 & 46.7 & 13.3 & 10.0 & 8.0\\
\bottomrule
\end{tabular}
\caption{One-to-many experiments on research paper dataset}
\label{tab:one-to-many-paper}
\end{table*}

\paragraph{Real-scenario Test}
This experiment demonstrated the remarkable efficacy of the DAS framework in academic author identification tasks. As shown in Table \ref{tab:real-scenario-paper}, DAS achieved 92\% accuracy on the Rank@20 metric with a mere 8\% missing rate, revealing that writing style-based author recognition could create substantial privacy vulnerabilities even in strictly anonymized academic scenarios. This finding carries significant implications for the academic publishing system, serving as a critical warning. Given the prevalent preprint culture in computer science, particularly in arXiv repositories, attackers could potentially compromise the double-blind review mechanism by analyzing authors' historical publications through DAS's sequential analysis.

\begin{table}[htbp]
\centering
\begin{adjustbox}{width=0.95\columnwidth}
\footnotesize 
\begin{tabular}{l*{5}{c}}
\toprule
& Rank@5 & Rank@10 & Rank@15 & Rank@20 & Miss \\
\midrule
DAS & 66\% & 80\% & 88\% & 92\% & 8\%   \\
\bottomrule
\end{tabular}
\end{adjustbox}
\caption{Real scenario experiments on research paper dataset}
\label{tab:real-scenario-paper}
\end{table}

\begin{table*}[htb]
\centering
\footnotesize
\begin{tabular}{l*{5}{c}}
\toprule
& Domain & \#Documents & \#Authors & Avg.Length & Source \\
\midrule
Review & Peer Review & 147,367 & N/A & 2,655 & OpenReview API \\
Blog & Personal Blogs & 15,000 & 1,500 & 116 & \citet{Schler2006EffectsOA} \\
Email & Corporate Communication & 8,700 & 174 & 197 & \citet{klimt2004enron} \\
Paper & Academic Writing & 26,632 & 1,500 & 7,383 & \citet{wen2024aidbench} \\
\bottomrule
\end{tabular}
\caption{Detailed characteristics of evaluation datasets}
\label{tab:dataset-details}
\end{table*}

 \section{Further explanation of random baseline}

To rigorously establish the theoretical baseline for random guessing performance, we provide a detailed derivation of the combinatorial probability model. Consider a candidate pool containing N documents where each target document has m same-author documents (not including the target itself). The probability of selecting at least one same-author document when randomly choosing k candidates follows hypergeometric distribution principles.

Let X be the random variable representing the number of same-author documents in a random sample of size k. The probability of selecting exactly s same-author documents is:

\begin{equation}
P(X = s) = \frac{\binom{m}{s}\binom{N-m}{k-s}}{\binom{N}{k}}.
\end{equation}

The probability of selecting at least one same-author document (Rank@k metric) is therefore:

\begin{equation}
\text{Rank@k} = 1 - P(X = 0) = 1 - \frac{\binom{N-m}{k}}{\binom{N}{k}}
\end{equation}

Where $\binom{N}{k}$ represents total ways to choose k documents from N and $\binom{N-m}{k}$ counts the ways to choose k documents excluding all m same-author documents.

For concrete illustration, we used a set of typical data for calculation: N=44770 (total documents), m=4 (same author documents), k=20 (top candidates).

The random baseline probability calculates as:
\begin{equation}
\text{Rank@20} = 1 - \frac{\binom{44764}{20}}{\binom{44770}{20}} \approx 0.014\%
\end{equation}

We ultimately aggregate the data from each year, where final result matches the 0.13\% random baseline reported in Table 1, demonstrating that DAS's 44\% Rank@20 accuracy represents a 338x improvement over random chance. The combinatorial approach provides exact theoretical expectations against which empirical results can be meaningfully compared.

\section{Dataset Details}

We conducted comprehensive evaluations across four key datasets representing distinct textual domains and privacy challenges:

\textbf{Blog} Curated from the Blog Authorship Corpus~\citep{Schler2006EffectsOA}, this dataset contains 15,000 posts by 1,500 active bloggers, capturing informal writing styles and personal expression patterns that enable authorship analysis of user-generated content.
\begin{table*}[htb]
\centering
\footnotesize 
\begin{tabular}{l*{6}{c}}
\toprule
\multirow{2}{*}{} & 
\multicolumn{3}{c}{2 Authors} & 
\multicolumn{3}{c}{5 Authors} \\
\cmidrule(lr){2-4} \cmidrule(lr){5-7}
& Rank@1 & Rank@3 & Rank@5 & Rank@1 & Rank@3 & Rank@5 \\
\midrule
DeepSeek-R1 & 96.3 & 100.0 & 100.0 & 85.2 & 88.9 & 96.3\\
Claude-3.5-Sonnet & 100.0 & 100.0 & 100.0 & 93.3 & 100.0 & 100.0\\
\midrule
& Prec@1 & Prec@3 & Prec@5 & Prec@1 & Prec@3 & Prec@5 \\
\midrule
DeepSeek-R1 & 96.3 & 97.5 & 90.4 & 85.2 & 76.5 & 75.6\\
Claude-3.5-Sonnet & 100.0 & 95.6 & 93.3 & 93.3 & 97.8 & 90.7\\
\bottomrule
\end{tabular}
\caption{Evaluation of different models one-to-many identification capability on research paper dataset}
\label{tab:extended}
\end{table*}

\textbf{ICLR Review} The ICLR Review dataset aggregates 147,367 anonymized peer reviews from 2023-2025 conferences via OpenReview API, simulating real-world deanonymization attacks on academic review systems through structured metadata and textual analysis.

\textbf{Enron Email} Derived from the Enron corpus~\citep{klimt2004enron}, the benchmark includes 8,700 professionally written emails from 174 executives, preserving linguistic fingerprints while removing sensitive headers to study corporate communication privacy risks.

\textbf{Research Paper} The academic writing dataset comprises 24,095 single-author machine learning papers from arXiv CS.LG (2019-2024)~\citep{wen2024aidbench}, filtered to include only authors with $\ge 10$ publications, exposing stylistic consistency challenges in scholarly communication.

Table \ref{tab:dataset-details} provides comprehensive statistics for all datasets used in our experiments. The table includes key characteristics such as document domains, corpus sizes, author counts, average text lengths, and original data sources. All datasets were preprocessed to remove personally identifiable information while preserving linguistic patterns. Text lengths represent word counts after standard cleaning (stopword removal, punctuation stripping).

\subsection{Sensitivity Analysis}

\begin{figure}[htbp]
    \centering
    \includegraphics[width=1\linewidth]{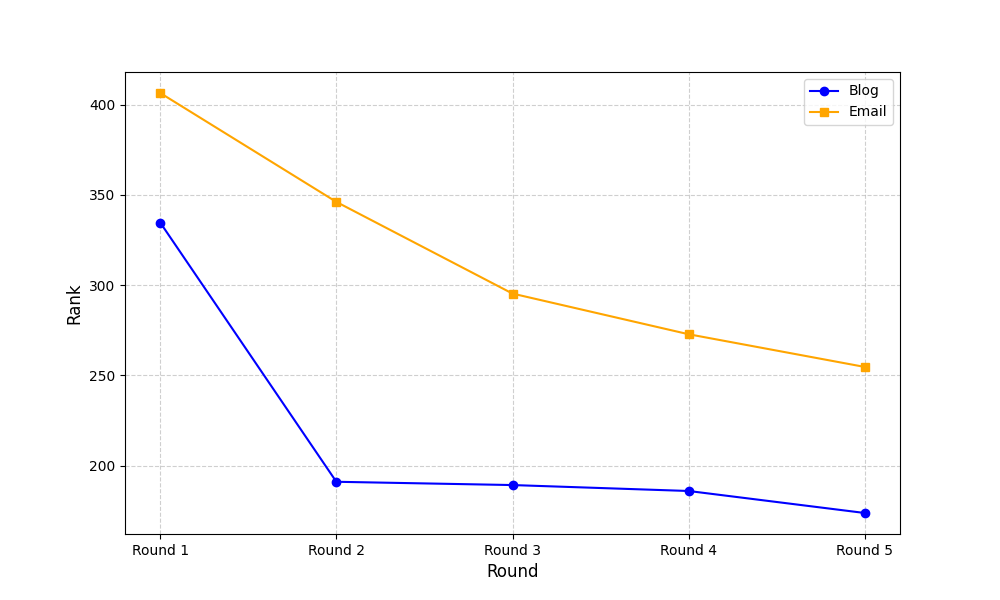}
    \caption{Mean rank of same author documents per round}
    \label{fig:sensitivity}
\end{figure}

Figure \ref{fig:sensitivity} reveals consistent performance improvement across text genres, with mean same-author rankings reduced by 47\% after five iterations. Despite the initial variance caused by text length and stylistic differences, DAS maintaind downward trajectory in both free-form blogs and structured emails. This demonstrates the framework's dual-phase synergy: coarse filtering eliminates stylistic outliers while fine-grained analysis identifies stable authorial patterns, enabling effective cross-domain discrimination. The convergence trend validates DAS's capacity to extract personalized expression signatures regardless of text structure.

\end{document}

%% file: introduction.tex
\section{Introduction}

Large language models (LLMs) have seen rapid and widespread adoption due to their remarkable ability to generate human-like text and follow complex instructions across domains \citep{zhao2023survey, chang2024survey, aw2023instruction}.  Such wide deployment in real-world systems, fundamentally reshapes how human interact with information. 

Alongside their impressive capabilities, however, LLMs also introduce new \emph{privacy concerns}. One emerging risk is the potential for LLMs to undermine anonymity in settings where privacy is paramount – for example, double-blind academic peer review\footnote{\url{https://openreview.net}}, whistleblower forums, or anonymous communication platforms. These systems depend on keeping authors’ identities hidden to protect fairness and safety, yet LLMs’ strong text analysis abilities may enable them to infer identifying signals, such as distinctive writing patterns or domain expertise, and thereby de-anonymize content intended to remain anonymous \citep{staab2023beyond, nyffenegger2023anonymity}

%While current privacy research on LLMs has primarily focused on models memorizing sensitive training data \citep{nasr2023scalable,carlini2021extracting}, far less attention has been paid to this stylometric threat: the ability of LLMs to attribute authorship to an “anonymous” text by comparing its linguistic patterns with other texts \citep{staab2023beyond, nyffenegger2023anonymity}. %If LLMs can effectively serve as universal writing-style fingerprints, they pose a serious challenge to privacy in online and institutional contexts that depend on author anonymity.

Existing work on authorship attribution offers limited guidance for this privacy risk. Traditional authorship attribution (AA) in NLP and forensic linguistics \citep{stamatatos2009survey, neal2017surveying, he2024authorship} is usually studied in a \textbf{closed-set} setting: a small, fixed list of candidate authors is given, and each author has labeled writing samples. This assumption underlies many stylometry benchmarks and PAN competitions \citep{bevendorff2022overview, stamatatos2018overview, bevendorff2023overview}, where the candidate pool is often only tens of authors and building per-author profiles or classifiers is feasible.

Real-world anonymous systems look very different: there may be tens of thousands of possible authors and no pre-labeled profile texts for any of them. In double-blind peer review, for example, the reviewer pool for major AI venues can be extremely large, yet we typically lack labeled writing from each reviewer. Authorship attribution at this scale, under such minimal assumptions, remains underexplored. Recent work \citet{Huang2024CanLL} prompts GPT-3/4 for author attribution on blogs and emails, it still considers relatively small candidate sets. In practice, de-anonymization may require searching tens of thousands of candidate texts with minimal supervision, a scenario where previous methods break down or become computationally infeasible, motivating new techniques for LLM-based de-anonymization at scale.

In this paper, we introduce \textbf{De-Anonymization at Scale (DAS)}, an LLM-based method for authorship matching in anonymous systems with tens of thousands of candidate texts. DAS uses a \emph{progressive elimination strategy}: we randomly partition candidate texts into small groups, prompt the LLM to select the most likely match to a query within each group, i.e., a one-to-many identification process, and iteratively re-group and re-compare the surviving candidates until we obtain a ranked top-k list.

To scale under a restricted LLM token budget, DAS employs a \emph{coarse-grained retrieval module} \citep{lewis2020retrieval} to shrink the search space. Given a query text, embedding-based retrieval narrows a corpus of up to 
$10^5$ candidate texts to, say, the top $10^3$. This retrieval step acts as a coarse but efficient sieve, ensuring that only the most likely candidates proceed to  the expensive LLM comparisons. This design makes large-scale de-anonymization computationally and economically feasible.

To further improve the accuracy, we employ a \emph{majority-voting-style scoring system} to enhance robustness with multiple independent runs. DAS repeats the  progressive selection with different random partitions, and each time  assigns scores to candidates that win comparisons. By aggregating these scores of  multiple runs, DAS produces the final ranking, favoring texts that consistently match the query across runs.

We evaluate DAS on both real-world and benchmark datasets. On anonymized double-blind peer-review data, DAS identifies same-author reviews from pools of thousands at rates well above chance, providing evidence that modern LLMs can threaten reviewer anonymity. On standard benchmarks, DAS also improves accuracy and scalability over direct LLM prompting, including the Enron email corpus \citep{klimt2004enron} and a large blogger dataset \citep{Schler2006EffectsOA}. Overall, our results show that LLM-enabled de-anonymization is a practical risk, motivating the development of stronger mitigation and privacy safeguards in the era of omnipresent large language models.

 The contributions are summarized as follows. 
\begin{itemize}
    \item 
    \textbf{New Privacy Risk.} We identify a realistic and  underappreciated privacy threat: by leveraging the power of modern LLMs, an adversary can de-anonymize texts in systems like double-blind peer review at rates significantly above chance. This finding urges a fundamental rethinking of how anonymous platforms are designed and secured against stylometric attacks.
    
\item \textbf{Technical Contribution.} We propose \textbf{De-Anonymization at Scale (DAS)}, a two-stage pipeline combining (1) a coarse dense-retrieval filter to narrow down tens of thousands of candidates to a short list and (2) an LLM-based sequential progression with a majority-voting scoring mechanism for fine-grained authorship attribution. This design enables both efficiency and accuracy in massive open-set authorship attribution scenarios.

    \item \textbf{Empirical Contribution.} Through extensive experiments on anonymized peer-review data and public benchmarks (Enron emails, large blogging corpora), we show that DAS substantially outperforms random guessing and existing LLM-based stylometric methods, achieving high accuracy and scalability in challenging, real-world deanonymization tasks.
\end{itemize}

Finally, despite the difficulty of attributing authorship in massive anonymized collections, our method is remarkably simple, further underscoring the severity of this risk.

%% file: related2.tex
\section{Related Work}
\label{sec:related-work}

\paragraph{Classical Authorship Attribution.} Authorship attribution (AA) has a long history in statistical stylometry. Traditionally, AA methods relied on human-defined linguistic features that capture an author's writing style \citep{holmes1994authorship,stamatatos2009survey}. Researchers engineered features such as character/word $n$-gram frequencies, vocabulary richness, function word usage, syntactic patterns, and other stylometric markers \citep{Seroussi2014}. These features, combined with machine learning classifiers (e.g., Bayesian or SVM-based techniques), proved effective on closed-world problems with small to medium author sets~\citep{madigan2005bayesian,koppel2014determining,koppel2007measuring,bevendorff2022overview}.  Comprehensive overviews of authorship attribution, including taxonomies of tasks (closed-set vs. open-set, verification vs. profiling) and feature categories, are provided by \citet{stamatatos2009survey,he2024authorship}.%For example, Madigan et al. \citep{Madigan2005} used a Bayesian logistic regression to attribute emails in the Enron corpus, and numerous studies benchmarked on datasets like the Enron emails \citep{Klimt2004} and blog posts \citep{Schler2006}. The Blog Authorship Corpus \citep{Schler2006} contains posts by nearly 20K bloggers and has been a popular testbed, as has the Enron corpus of about 160 authors' emails \citep{Klimt2004}. The PAN evaluation competitions have further spurred the development of standardized benchmark tasks and datasets for authorship analysis, including attribution, verification, and profiling across genres \citep{Bevendorff2020}. Notably, Koppel et al. demonstrated that attribution can scale to very large candidate sets, showing promising results even with thousands of possible authors \citep{Koppel2006}. They later highlighted the challenges of authorship attribution "in the wild," stressing issues like unseen authors and topic differences in real-world scenarios \citep{Koppel2011}. In forensic and security contexts, researchers have also treated authorship verification as a fundamental sub-problem: determining if two documents share an author \citep{Koppel2014}. This formulation underpins methods that can handle open-set or large-$N$ attribution by focusing on one-to-one stylistic similarity judgments.

\paragraph{Authorship Attribution with Large Language Models.} The recent generation of large language models (LLMs) has opened new avenues for authorship analysis. Unlike fixed feature extractors, LLMs can be prompted to perform complex NLP tasks zero- or few-shot, without fine-tuning on task-specific data. Initial studies have started to evaluate LLMs on authorship attribution and verification. \citet{Hung2023WhoWI} prompt LLMs to produce explanations for  authorship verification. \citet{Huang2024CanLL} systematically tested GPT-3.5 and GPT-4 on both verification and closed-set attribution with up to tens of candidates, where zero-shot GPT-based models can match or even surpass fine-tuned BERT classifiers on certain datasets. By employing specially crafted prompts that encourage the LLM to explain its decisions, they also extract human-interpretable justifications for the model’s predictions. \citet{gorovaia2024sui} similarly showed that an LLM (GPT-3) could robustly verify authorship of Latin texts in a zero-shot manner. Overall, these works suggest LLMs hold promise for authorship tasks, especially when fine-tuning data is scarce. 

\paragraph{De-anonymization, Privacy, and Adversarial Stylometry.} Authorship attribution techniques pose dual-use concerns: the same tools that identify authors can undermine anonymity and privacy. The security community has long studied ``stylometric attacks'', where an adversary de-anonymizes an author by matching their writing style across anonymous texts. A landmark study \citep{koppel2006authorship} illustrated that distinguishing tens of thousands of authors is theoretically feasible given sufficient text%, raising alarms about anonymity breaches
. Subsequent work demonstrated the practicality of such attacks and also how authors might evade them~\citep{brennan2012adversarial,emmery2021adversarial}. %Brennan et al. \citep{Brennan2012} showed that even simple manual obfuscation (e.g., deliberately changing one's writing style) can dramatically degrade attribution accuracy, while more sophisticated approaches use machine translation or automated word substitutions to mask telltale stylistic markers \citep{Emmery2021}. 
At the same time, researchers have developed methods to detect when text has been manipulated or when a writing style is inconsistent ~\citep{bevendorff2024overview}.  %On the offensive side, large-scale studies have applied authorship attribution to high-profile anonymous cases: Cafiero and Camps \citep{Cafiero2023} used a supervised stylometric classifier to analyze the QAnon conspiracy writings, successfully narrowing down candidate identities. 

The broader implications of such de-anonymization capabilities have been highlighted in survey \citep{huang2025authorship}, which call for more research on privacy-preserving text rewriting and adversarial robustness in authorship analysis. % These works situate the current trend of LLM-based attribution in the context of decades of prior research, underlining how the fundamental challenges of genre effect, topic influence, and data scarcity persist even as new models and methods emerge. and a recent survey by  \citet{huang2025authorship}